\RequirePackage{fix-cm}
\documentclass[smallextended]{svjour3}       
\smartqed  
\usepackage{graphicx}
\usepackage{amssymb,amsmath,latexsym}
\usepackage{mathtools,multirow,color,footmisc,multicol,hhline,float,varwidth,array}
\usepackage{longtable}
\usepackage{tipa}
\usepackage{tipx}
 \journalname{Cryptography and Communications}
\begin{document}
\newcommand\myeq{\stackrel{\mathclap{\normalfont\mbox{\tiny{def}}}}{=}}
\newcommand{\FF}{\mbox{$\mathbb{F}$}}
\newcommand{\m}{\mbox{$\mathbf{m}$}}
\newcommand{\e}{\mbox{$\mathbf{e}$}}
\newcommand{\ob}{\mbox{$\overline{\omega}$}}
\newcommand{\om}{\mbox{$\omega$}}
\newcommand{\Tr}{\mbox{Tr}}
\newcommand{\C}{\mbox{$\cal C$}}
\newcommand{\Cperb}{\mbox{$\C^\bot$}}
\newcommand{\ben}{\begin{equation*}}
\newcommand{\een}{\end{equation*}}
\renewcommand{\thefootnote}{\fnsymbol{footnote}}

\title{McNie: A new code-based public-key cryptosystem
}


\author{Jon-Lark Kim   \and  Young-Sik Kim  \and Lucky Galvez  \and  Myeong Jae Kim   \and  Nari Lee}


\institute{J.L. Kim, L. Galvez, M.J. Kim, N. Lee \at
             Department of Mathematics, Sogang University, Seoul 04107, South Korea       \\
             \email{ jlkim@sogang.ac.kr, legalvez97@gmail.com,
device89@snu.ac.kr, narilee3@gmail.com}           
           \and
           Y.S. Kim \at
              Department of Information and Communication Engineering,
Chosun University, Gwangju 61452, South Korea \\
\email{ iamyskim@chosun.ac.kr}
}

\date{Received: date / Accepted: date}

\maketitle

\begin{abstract}
In this paper, we suggest  a new code-based public key encryption scheme, called McNie.  McNie is a hybrid version of the McEliece and Niederreiter cryptosystems and its security is reduced to the hard problem of syndrome decoding. The public key involves a random generator matrix which is also used to mask the code used in the secret key. This makes the system safer against known structural attacks. In particular, we apply rank-metric codes to McNie. 
\end{abstract}

\section{Introduction}\label{sec:intro}
Code-based cryptography  was introduced in 1978 by McEliece using binary Goppa codes, providing efficient encryption and decryption algorithms, and   reasonably good security level \cite{McE}. The security of the McEliece cryptosystem relies on the hardness of decoding a random  linear code in some metric.   
Since the key size  in the original proposal by McEliece is significantly larger than those of RSA and ECC, it has been considered inefficient.  To overcome this problem, there has  been tremendous progress in  code-based cryptography using various codes replacing Goppa codes \cite{Bal,BL,Can,JM,LJ,Mis,Mon,Nie,Sid}.
However  most of the proposed cryptosystems are broken by structural attacks using the algebraic structures of the codes used \cite{BC,CB,CMP,FM,LT,MS,Sen,SS,Wie}.

In 1991, Gabidulin, Paramonov and Tretjakov (GPT) \cite{Gabi3}  proposed  a variant of McEliece based on Gabidulin codes, which are a family of the so-called rank-metric codes with an efficient algebraic polynomial time decoding algorithm. The GPT cryptosystem, unfortunately, was broken by Overbeck \cite{Over} using the invariant property of a vector space under Frobenius automorphism. Afterwards a new family of rank-metric codes was introduced by Gaborit, Ruatta, Schrek and Zemor \cite{Gabo1}, namely Low Rank Parity Check (LRPC) codes. So far, the double-circulant LRPC (DC-LRPC) code based cryptosystem   is known to be secure against the existing attacks.

We propose a new code-based public key cryptosystem, which we call \textit{McNie}, as a candidate for the NIST Post-quantum Cryptography Standardization \cite{NIST}.  McNie is a hybrid version of the McEliece and Niederreiter cryptosystems. Unlike the McEliece cryptosystem, we employ a random code in the encryption so that McNie is secure against structural and information set decoding attacks. We proved that it is not more difficult to break McEliece than McNie. 

 After the submission, we have received two attacks. First, Gaborit \cite{PQCcomment} proposed a message recovery attack to lower the security level of the original submitted parameters. Then, Lau and Tan \cite{Lau} proposed a key-recovery attack. In this paper, we describe McNie and show how to avoid these attacks. Furthermore, the McNie document submitted to NIST did not describe the security reduction. So we show that the security of McNie is based on the (Rank) Syndrome Decoding Problem. 

The remainder of this paper is organized as follows: In Section \ref{sec:Prel}, we introduce  basic notions and terminologies used throughout the paper and a modified decoding algorithm for LRPC codes reducing the  complexity in \cite{Gabo1}.  In Section \ref{sec:KeyGen}, we present the general key generation, encryption and decryption algorithms for McNie. We then show how key sizes can be reduced using 3-quasi or 4-quasi cyclic LRPC codes. In Section \ref{Sec:SecArg}, we analyze the security of the proposed cryptosystem against known attacks and examine the parameter restrictions. In Section \ref{Sec:Para}, we suggest parameters for 128-bit, 192-bit and 256-bit security levels, and provide the corresponding key sizes,  comparing with other code-based public key sizes.

This work was partially presented at the First NIST Post-quantum Cryptography Conference in April 2018, but it was never published as a conference or journal paper.

\section{Background}\label{sec:Prel}

\subsection{Rank codes}

We begin by defining the rank metric codes or simply, rank codes. Essentially, rank  codes are linear codes equipped with the rank metric, instead of the usual Hamming metric. There are two representations of rank codes, as we will see later, which  are actually related. One of them was first introduced by Delsarte in  the $matrix$ $representation$  originally as a bilinear form \cite{Del}. The other,  the $vector$ $representation$,  was  introduced by Gabidulin in his seminal paper \cite{Gabi1}.

\begin{definition}
{\em $Rank$ $codes$  in $matrix$ $representation$ are subsets of the normed space $\{ \FF_q^{m \times n}, Rk \}$ of $m\times n$ matrices over a finite field $\FF_q$, where the norm of a matrix $M \in \FF_q^{m \times n}$ is defined to be its algebraic rank $Rk(M)$ over $\FF_q$. The $rank$ $distance$ $d_R(M_1,M_2)$ between two matrices $M_1$ and $M_2$ is the rank of their difference, i.e., $d_R(M_1,M_2) =Rk(M_1-M_2)$. The $rank$ $distance$ of a matrix rank code $\mathcal{M}\subset \FF_q^{m \times n}$ is defined as the minimal pairwise distance:
\[d(\mathcal{M})=d=\min(Rk(M_i - M_j) : M_i, M_j \in \mathcal{M}, i\neq j).\] }
\end{definition}

Notice that for a given ordered basis $\beta = \{ \beta_1, \beta_2, \ldots, \beta_m \}$ of $\FF_{q^m}$ over $\FF_q$, every vector $\mathbf{v}=(v_1,v_2,\ldots,v_n)  \in \FF_{q^m}^n$ corresponds to a matrix $\bar{\mathbf{v}}$ whose $i^{th}$ column consists of the coefficients when $v_i$ is written in terms of the basis $\beta$. Moreover, $Rk(\mathbf{v})=Rk(\bar{\mathbf{v}})$ and this is independent of the chosen basis. Therefore, every rank code $C$ in vector representation can be expressed as a rank code in matrix representation, with respect to the basis $\beta$. Throughout the rest of this paper, all rank  codes being considered are in vector representation.

Another difference between rank codes and codes in the Hamming metric is the definition of the support of a codeword.

\begin{definition}\label{def3}
{\em Let $\mathbf{x} = (x_1, x_2, \ldots, x_n) \in \FF^n_{q^m}$ be a vector of rank $r$. We denote by $E$ the $\FF_q$-sub vector space of $\FF_{q^m}$ generated by the entries of $\mathbf{x}$, i.e.,  $E= \left< x_1, x_2, \ldots, x_n \right>$. The vector space $E$
is called the $support$ of $\mathbf{x}$.}
\end{definition}

Many well-known codes in the Hamming metric have rank metric analogues. Of particular interest in many cryptographic applications are the so-called LRPC codes, which can be considered as the analogue of LDPC codes in the rank metric. 

\begin{definition}
{\em A $Low$ $Rank$ $Parity$ $Check$ (LRPC) code of rank $d$, length $n$ and dimension $k$ over $\FF_{q^m}$ is a code that has for its parity check matrix an $(n-k)\times n$ matrix $H=(h_{ij})$ such that the sub-vector space of $\FF_{q^m}$ generated by its coefficients $h_{ij}$ has dimension at most $d$. We call this dimension the weight of $H$. Letting $\mathcal{F}$ be the sub-vector space of $\FF_{q^m}$ generated by the coefficients $h_{ij}$ of $H$, we denote one of its bases by $\{F_1, F_2, \ldots, F_d\}$.}
\end{definition}
Simply put, the entries of the parity check matrix $H$ belong to a subfield $\mathcal{F}$ of small dimension and can be expressed as $h_{ij} = \sum_{l=1}^d h_{ijl} F_l$, for some $h_{ijl} \in \FF_q$ and for any $1 \leq i \leq n -k$, $1 \leq j \leq n$.
\medskip

 A special type of LRPC code is the following:

\begin{definition}
{\em An LRPC code such that its parity check matrix $H$ is a  quasi-cyclic matrix of low weight $d$ is called a $quasi$-$cyclic$ LRPC (QC-LRPC) code. If $H$ is double-circulant, i.e., the concatenation of two cyclic matrices, then the code is called a $double$-$circulant$ LRPC (DC-LRPC) code.}
\end{definition}

\subsection{Rank code based cryptography}\label{RSD}
In 1978, McEliece proposed a public key cryptosystem using a family of error-correcting codes with efficient decoding algorithm \cite{McE}. In the original construction, a  binary $t$-error-correcting Goppa code with generator matrix $G$ is generated together with an invertible matrix $S$, called the scrambler matrix, and a permutation matrix $P$. The triple $(G,S,P)$ forms the private key. The matrices $S$ and $P$ are used to hide the algebraic structure of $G$ so the public key is $G_{pub} = SGP$. The message $\m$ is then encrypted as $c=\m G_{pub} + \e$ where $\e$ is a random vector of weight at most $t$. In order to decrypt the received message $c$, note that $cP^{-1} = \m SG+\e P^{-1}$. Since $\e P^{-1}$ is still of weight less than $t$, we can apply the decoding algorithm to obtain $\m S$ and finally multiplying by $S^{-1}$ to obtain $\m$.

One drawback of the McEliece cryptosystem is the large key size. In recent years, rank codes have also been used in order to reduce key sizes. For instance, it is proposed in \cite{Gabo1} that LRPC codes are very fitting to use in a McEliece-type rank code based cryptosystem.  This is so because LRPC code has an efficient decoding algorithm, as shown in the next section. Moreover, in the LRPC cryptosystem, the private key is the low rank parity check matrix $H$ and the public key is the generator matrix $G$, hence eliminating the need for a scrambler matrix and a permutation matrix. This is the first rank metric based cryptosystem with poor algebraic structure, and is still considered secure. Furthermore, the double circulant variant provides very small key sizes.

Like many cryptographic systems, the McEliece cryptosystem is based on the hardness of a certain problem known as the syndrome decoding problem. This problem is generalized in the case of rank metric as follows:
\medskip

\noindent \textbf{Rank Syndrome Decoding (RSD) Problem }

\noindent Let $H$ be a $(n - k) \times n$ matrix over $\FF_{q^m}$ with $k \leq n$, $\mathbf{s} \in \FF^k_{q^m}$ and $r$  an integer. Find $\mathbf{x}$ such that $Rk(\mathbf{x}) = r$ and $H\mathbf{x}^t = \mathbf{s}$.
\medskip

The RSD problem is proven to be probabilistically NP-hard \cite{Gaborit2}.  

\subsection{Decoding LRPC codes}

 The decoding algorithm for LRPC codes was introduced in \cite{Gabo1}. The idea of the decoding algorithm is to recover the support $E$ of the error from the space $S$ generated by the entries of the syndrome and the known small basis $\mathcal{F}$ of $H$. If the error space $E$ is obtained, it is easy to recover the exact coordinates of the error vector $\e$ by solving a linear system.

As in the definition of LRPC code,  let $H=(h_{ij})\subseteq \FF_{q^m}^n$ such that each $h_{ij}$ belongs to a subspace $\mathcal{F}$ with basis $\left\{F_{1},F_{2},\ldots,F_{d}\right\} $ and write $h_{ij}= \sum_{v=1}^{d}h_{ijv}F_{v}$. Suppose that the received word is  $\mathbf{y}=\m G+ \e \in \FF_{q^m}^n$ where $\e=(e_1,\ldots,e_n) $ is the error vector of rank $r$ lying in the subspace $E$ with basis $\left\{ E_{1},E_{2},\ldots,E_{r}\right\} $.  Rewrite the entries of $\e=(e_{1},e_{2},\ldots,e_{n})$ in terms of the basis $\left\{ E_{1},E_{2},\ldots,E_{r}\right\} $, i.e., \[e_{j}={\displaystyle \sum_{u=1}^{r}e_{ju}E_{u}}.\]
Let $\e^\prime =(e_{11},e_{12},\ldots,e_{nr}).$
Denote by $\left\langle \mathcal{F}.E\right\rangle $ the product space generated by $\left\{ F_{1}E_{1},F_{1}E_{2},\ldots, \right.$ $\left. F_{d}E_{r}\right\} $ and suppose the dimension of $\left\langle \mathcal{F}.E\right\rangle $ is $rd.$

We will transform the syndrome equation $H\e^{t}=\mathbf{s}$ as an equation over the base field $\FF_{q}$. First, rewrite the entries of $\mathbf{s}=(s_{1},s_{2},\ldots,s_{n-k})$ in terms of the basis $\left\{ F_{1}E_{1}, F_{1}E_{2}, \ldots,F_{d}E_{r}\right\} $ for $\mathcal{F}.E$,  i.e., \[s_{i}={\displaystyle \sum_{v=1}^{d}\sum_{u=1}^{r}s_{ivu}F_{v}E_{u}}.\]  So we have $\mathbf{s}^\prime=(s_{111},s_{112},\ldots,s_{(n-k)dr})^{t}$, the representation of $\mathbf{s}$ with entries in $\FF_{q}$.
So the syndrome equation is equivalent to \[ \sum_{j=1}^{n}\sum_{v=1}^{d}\sum_{u=1}^{r}h_{ijv}e_{ju}F_{v}E_{u}=\sum_{u=1}^{r}\sum_{v=1}^{d}s_{ivu}F_{v}E_{u}\] written over $\FF_q$. Equating coefficients of $F_{v}E_{u}$ ($1\leq v\leq d$, $1\leq u\leq r$) we have $\sum_{j=1}^{r}h_{ijv}e_{ju}=s_{ivu}$.

This means that there exists a $(n-k)rd\times nr$ matrix $A_{H}^{r}$ such that $A_{H}^{r}\e^{\prime t}=\mathbf{s}^\prime$. Note that the entries of $A_{H}^{r}$ are the $h_{ijv}$'s and some $0$'s. In fact, if $A_{H}^{r} = (a_{ij})$, then \[a_{u+(v-1)r+(i-1)rd,~u+(j-1)r}=h_{ijv}\] for $1\leq v\leq d$, $1\leq u\leq r$, $1\leq i\leq n-k$, $1\leq j\leq n$ and $0$'s elsewhere. The matrix will appear as follows:

\[ A_{H}^{r} = \left [
\begin{array}{ccccccccccc}
h_{111}&&&h_{121}&&&&& h_{1n1}&&\\
&\ddots&&&\ddots & &\cdots &&&\ddots&\\
&&h_{111}&&&h_{121}&&&&&h_{1n1}\\
h_{112}&&&h_{122}&&&&&h_{1n2}&&\\
&\ddots&&&\ddots&&&\cdots&&\ddots&\\
&&h_{112}&&&h_{122}&&&&&h_{1n2}\\
& &&&&&&&&&\\
& \vdots&&&\vdots&&\cdots&&&\vdots&\\
& &&&&&&&&&\\
h_{(n-k)1d}&&&h_{(n-k)2d}&&&&&h_{(n-k)nd}&&\\
&\ddots&&&\ddots&&&\cdots&&\ddots&\\
&&h_{(n-k)1d}&&&h_{(n-k)2d}&&&&&h_{(n-k)nd}\\
\end{array}
\right]
\]

\medskip

\begin{definition}
Let $A_H$ be an $nr \times nr$ invertible submatrix of $A^r_H$. The matrix $D_H = A^{-1}_H$ is called a decoding matrix of $H$.
\end{definition}

\noindent The decoding algorithm can be summarized in the following steps:
\begin{enumerate}
\item Compute the syndrome vector $H\mathbf{y}^t = \mathbf{s} = (s_1,\ldots,s_{n-k})$ and the syndrome space $S= < s_1,\ldots,s_{n-k}>$.
\item Define $S_i = F^{-1}_i S$, the subspace where all generators of $S$ are multiplied by $F^{-1}_i$. Compute the support of the error $E = S_1 \cap S_2 \cap \cdots \cap S_d$, and compute a basis $\{ E_1, E_2, \ldots, E_r \}$
of $E$.
\item Solve the system $H\e^t = \mathbf{s}$, where the equations $H\e^t$ and the syndrome coordinates $s_i$ are written as elements of the product space $P =< \mathcal{F}.E >$ in the basis $\{F_1E_1, \ldots, F_1E_r,$  $\ldots, F_dE_1, \ldots, F_dE_r \}$. The
system has $nr$ unknowns (the $e_{ij}$ ) in $\FF_q$ and $(n-k)rd$ equations from the syndrome. The decoding matrix $D_H$ permits to recover directly the $nr$ values $e_{ij}$ from $nr$ positions of the $s_i$ written in product basis by a simple multiplication.
\item Recover $\m$ from the system $\m G = \textbf{y}- \e$.
\end{enumerate}

The following theorem from \cite{Gabo1} gives the error-correction failure probability and decoding complexity of the above algorithm.

\begin{theorem}\label{LRPCcomplex}
Let $H$ be a $(n-k)\times n$ dual matrix of an LRPC code with low rank $d \geq 2$ over $\FF_{q^m}$. Then the decoding algorithm given above decodes a random error $\e$ of low rank $r$ such that $rd \leq n-k$, with failure probability $q^{-(n-k+1-rd)}$ and complexity $r^2(4d^2m + n^2)$.
\end{theorem}

\subsection{Reducing the decoding complexity by a factor of the number $r$ of errors}
The decoding algorithm proposed by Gaborit et al.  \cite{Gabo1} is proven fast and efficient, performing in polynomial time. However the matrix $A_{H}^{r}$ used in the decoding has many pointless zeros and this leads to the increment of decoding complexity.

Here we introduce a modified decoding algorithm, reducing the complexity from $r^2(4d^2m+n^2)$ to $r(4d^2m+n^2)$ by a factor of $r$. The decoding complexity can be lowered by modifying the matrix $A_{H}^{r}$ to $K_{H}^{r}=(k_{ij})$, where $k_{i+(n-k)(v-1),j}=h_{ijv}$ for  $1\leq v\leq d$, $1\leq u\leq r$, $1\leq i\leq n-k$, and $1\leq j\leq n$. Note that $K_{H}^{r}$ is now then an $(n-k)d\times n$ matrix.

{\small
\[ K_{H}^{r}= \left [
\begin{array}{cccc}
h_{111}&h_{121}&\cdots& h_{1n1}\\
h_{211}&h_{221}&\cdots& h_{2n1}\\
\vdots&\vdots&&\vdots\\
h_{(n-k)11}&h_{(n-k)21}&\cdots&h_{(n-k)n1}\\
h_{112}&h_{122}&\cdots& h_{1n2}\\
h_{212}&h_{222}&\cdots& h_{2n2}\\
\vdots&\vdots&&\vdots\\
h_{(n-k)12}&h_{(n-k)22}&\cdots&h_{(n-k)n2}\\
&&&\\
\vdots&\vdots&&\vdots\\
&&&\\
h_{11d}&h_{12d}&\cdots& h_{1nd}\\
h_{21d}&h_{22d}&\cdots& h_{2nd}\\
\vdots&\vdots&&\vdots\\
h_{(n-k)1d}&h_{(n-k)2d}&\cdots&h_{(n-k)nd}\\
\end{array}
\right]
\]
}

\begin{definition}
Let $K_H$ be an $n \times n$ invertible submatrix of $K^r_H$. The matrix $D_H^\prime = K^{-1}_H$ is called a $K$-decoding matrix of $H$.
\end{definition}

We can write  the error vector $\e$ and the syndrome $\mathbf{s}$ as follows:

{\footnotesize
\[  \e=
\left [
\begin{array}{cccc}
e_{11}&e_{12}&\cdots& e_{1r}\\
e_{21}&e_{22}&\cdots& e_{2r}\\
\vdots&\vdots&&\vdots \\
e_{n1}&e_{n2}&\cdots&e_{nr}\\
\end{array}
\right],~~~ ~~\mathbf{s}=
\left [
\begin{array}{cccc}
s_{111}&s_{112}&\cdots& s_{11r}\\
s_{211}&s_{212}&\cdots& s_{21r}\\
\vdots&\vdots&&\vdots \\
s_{(n-k)11}&s_{(n-k)12}&\cdots&s_{(n-k)1r}\\
s_{121}&s_{122}&\cdots& s_{12r}\\
s_{221}&s_{222}&\cdots& s_{22r}\\
\vdots&\vdots&&\vdots \\
s_{(n-k)21}&s_{(n-k)22}&\cdots&s_{(n-k)2r}\\
&&&\\
\vdots&\vdots&&\vdots\\
&&&\\
s_{1d1}&s_{1d2}&\cdots& s_{1dr}\\
s_{2d1}&s_{2d2}&\cdots& s_{2dr}\\
\vdots&\vdots&&\vdots \\
s_{(n-k)d1}&s_{(n-k)d2}&\cdots&s_{(n-k)dr}\\
\end{array}
\right]
\]
}

\section{The McNie Public-key Encryption} \label{sec:KeyGen}

\subsection{Design of the scheme}
The following describes the general key generation, encryption and decryption algorithms of the proposed McNie public key encryption scheme. Notice that in this general scheme, a random code is used as part of the public key, which is not related to the code used in the secret key. 
\medskip

\noindent\fbox{ \parbox{0.965\textwidth}{
\textbf{Key generation}

Generate a parity check matrix $H \in \FF_{q^m}^{(n-k) \times n}$ of an $r$-error-correcting code with an efficient decoding algorithm $\Phi_H$, an invertible matrix $S \in \FF_{q^m}^{(n-k) \times (n-k)}$ and an $n \times n$ isometry $P$.

Generate a generator matrix $G^\prime$ of a random $[n,l]$ code over $\FF_{q^m}$, where $l > n-k$ and compute  $F=G^\prime P^{-1}H^TS$.
\begin{itemize}
\item\textbf{Private key: $(S,H,P)$ }
\item\textbf{Public key: $(G^\prime,F)$}
\end{itemize}
\smallskip

\noindent\textbf{Encryption}

To encrypt a plaintext $\m \in \FF_{q^m}^l$, generate a random vector $\e \in \FF_{q^m}^n$ of weight at most $r$. Compute $\mathbf{c_1} = \m G^\prime + \e$ and $\mathbf{c_2} = \m F$.
\begin{itemize}
\item \textbf{Ciphertext: $(\mathbf{c_1},\mathbf{c_2})$}
\end{itemize}
\smallskip

\noindent\textbf{Decryption}

To decrypt a received vector $\mathbf{y}=(\mathbf{c_1},\mathbf{c_2})$, compute $\mathbf{s^\prime} = \mathbf{c_1}P^{-1}H^T-\mathbf{c_2}S^{-1}=\e P^{-1}H^T$, then $\e P^{-1} = \Phi_H(s^\prime)$. Apply $P$ to obtain $\e$ and recover $\m$ from $\mathbf{c_1}-\e = \m G^\prime$ using linear algebra.

}}
\bigskip

Note that in the decryption step, we use the decryption algorithm for $H$ even though $\mathbf{c_1}$ is encrypted using $G^\prime$ and so the weight of the error vector $\e$ depends on $H$ and not on the public key $G^\prime$. Since $G^\prime$ is random and not related to $H$ in any way, the error vector $\e$ can have weight beyond the error-correction capability of the code generated by $G^\prime$. In other words, if the error-correction capability of the code generated by $G^\prime$ is less that $r$, decryption is still possible but an attacker would face a more difficult task of decoding the message using $G^\prime$. This provides an added security compared to the usual McEliece setting.

\begin{remark}
We note that McNie is a general scheme of McEliece. Let $G$ be a generator matrix of the code with an $(n-k)\times n$ parity check matrix $H$. Suppose that the public matrix is $G^\prime = S'GP$, where $S$ is an invertible matrix and $P$ is a permutation matrix. Then $F=(S'GP)P^{-1}H^TS= \mathbf{0} \in \FF_{q^m}^{k\times (n-k)}$ since $GH^T=\mathbf{0}$ so $\mathbf{c_2} = \mathbf{0} \in \FF_{q^m}^{n-k}$ and $\mathbf{c_1}=\m G^\prime+\e=\m S'GP+\e$ is the ciphertext for the McEliece cryptosystem. 
\end{remark}

\begin{remark} Notice that the dimension $l$ of a public key $G^\prime$ should be greater than the dimension $n-k$ of the parity check matrix $H$. Otherwise, the attacker may recover the message vector from  the ciphertext $\mathbf{c_2} = \m F$, where $\m =(m_1, m_2, \ldots, m_l)$  and  $F$ an $l \times (n-k)$ matrix.  That is, if $l\leq rank(F)$, the linear system with $l$ unknowns will give us the unique solution,  which is the message $\m$.  When $rank(F) < l$, there are $q^{m(l-rank(F))}$ possible solutions for $\m$. 

\end{remark}

\begin{remark} Suppose $F=G^\prime P^{-1}H^TS= G^\prime H_0$ for some $n \times (n-k)$ matrix $H_0$ over $\FF_{q^m}$, then the number of possible choices for $H_0$ is $q^{m(n-l)(n-k)}$. 

\end{remark}

McNie is designed to use the parity check matrix $H$ of any $r$-error correcting code belonging to a family of codes with known efficient decoding algorithm. In the following sections, we use rank metric codes, particularly the quasi-cyclic variants of LRPC codes in order to reduce key sizes.
\subsection{Generating keys using $3$-quasi-cyclic LRPC codes}
Lau and Chan \cite{Lau} gave a structural attack for these codes when the isometric matrix $P$ is the identity matrix. We can avoid this by choosing a non-identity matrix $P$ which is given in detail as follows.

The size of public key $(G^\prime, F)$ can be reduced by using quasi-cyclic codes.  Let $G^\prime$ be an $[n,\frac{2n}{3}]$  be a generator matrix for a random 3-quasi-cyclic code over $\FF_{q^m}$ and $H$ be a parity check matrix for an  $[n,\frac{n}{3}]$  3-quasi-cyclic LRPC code over $\FF_{q^m}$. Then  $G^\prime$ and $H$ will be in the  following form:

\[
G^\prime=\left[\begin{array}{ccc}
I_{\frac{n}{3}} & 0 & G_{1}\\
0& I_{\frac{n}{3}} & G_{2}
\end{array}\right],
\quad
H=\left[\begin{array}{ccc}
H_{1} & H_{2} & H_{3}
\end{array}\right]
\]
where $G_{i}$ and  $H_{j}$  are circulant square matrices of size $\frac{n}{3}$ for $1\leq i\leq 2$ and $1\leq j \leq 3$.

Next, randomly generate an $\frac{n}{3}\times \frac{n}{3}$ nonsingular block cyclic matrix $S$ over $\FF_{q^m}$, and a nonsingular matrix $P$ over the base field $\FF_q$ such that  $P^{-1}=\left[\begin{array}{ccc}P_1 & P_2 & P_3 \\ P_4 & P_5 & P_6 \\ P_7 & P_8 & P_9 \end{array} \right]$ where each $P_i$ is a cyclic matrix.  Then  $F=G^\prime P^{-1}H^{T}S$ is

\begin{eqnarray*}
F & = & 
\left[\begin{array}{ccc}
I_{\frac{n}{3}} & 0 & G_{1}\\
0& I_{\frac{n}{3}} & G_{2}
\end{array}\right]
\left[\begin{array}{ccc}P_1 & P_2 & P_3 \\ P_4 & P_5 & P_6 \\ P_7 & P_8 & P_9 \end{array} \right]
\left[\begin{array}{c}
H_{1}^T\\
H_{2}^T\\
 H_{3}^T
\end{array}\right]S \\
& = & \left[\begin{array}{c}
(P_1+G_1P_7)H_1^T+(P_2+G_1P_8)H_2^T+(P_3+G_1P_9)H_3^T\\
(P_4+G_2P_7)H_1^T+(P_5+G_2P_8)H_2^T+(P_6+G_2P_9)H_3^T
\end{array}
\right]S,
\end{eqnarray*}
an $\frac{2n}{3} \times  \frac{n}{3}$ cyclic matrix over $\FF_{q^m}$.

Note that since $S$ is also block cyclic, it allows  a square matrix $F'$ of size $\frac{n}{3}$ to be cyclic in the following row echelon form of $F$:
\[
F=\left[\begin{array}{c}
I_{\frac{n}{3}}\\
F'
\end{array}
\right].
\]

\noindent\textbf{Reduced key size}

Since we can recover the rest of the rows of circulant matrices from the first row, we store only $\frac{2nm}{3} Log (q)$ bits for $G^\prime$ and $\frac{nm}{3}Log (q)$ bits for $F$ (instead of storing all the bits of the whole matrices) and so the total public key size is $nm Log (q)$. Here, $Log$ denotes the logarithm base 2.

\subsection{Generating keys using $4$-quasi-cyclic LRPC codes}\label{4quasi}
Lau and Chan \cite{Lau} gave a structural attack for these codes when the isometric matrix $P$ is the identity matrix. We can avoid this by choosing a non-identity matrix $P$ which is given in detail as follows.

In order to reduce key sizes in our system, we use circulant matrices and construct quasi-cyclic codes in the public and private keys.

Suppose $j=n/4$. Let $C$ be an $[n,2j]$ LRPC code over $\FF_{q^m}$ with rank $d$ and parity check matrix $H$ in the following form:
\begin{equation}\label{4quasiH}
H=\left[\begin{array}{cccc}
H_{1} & H_{2} & H_{3} & H_{4} \\
H_{5} & H_{6} & H_{7} & H_{8} \end{array}\right]
\end{equation}
where $H_{i}$, are circulant matrices of size $j\times j$ for  $i=1,2,\ldots,8$. Choose the nonsingular $2j \times 2j$ scrambler matrix $S$ to be block-circulant matrix, as well. Finally, take $P=\left[ \begin{array}{cccc} P_1 & P_2 & P_3 & P_4 \\ P_5 & P_6 & P_7 & P_8 \\ P_9 & P_{10} & P_{11} & P_{12} \\  P_{13} & P_{14} & P_{15} & P_{16} \end{array} \right]$ to be an invertible matrix over the base field $\FF_q$ where each $P_i$ is a $j \times j$ circulant matrix. The private key is $(H,S)$.

Let $G^{\prime}$ take the following form:
\begin{equation}\label{4quasiG}
G^{\prime}=\left[\begin{array}{cccc}
I_{j} & 0_{j} & 0_{j} & G_{1}\\
0_{j} & I_{j} & 0_{j} & G_{2}\\
0_{j} & 0_{j} & I_{j} & G_{3}
\end{array}\right]
\end{equation}
 where $I_{j}$ is the $j\times j$ identity matrix, $0_{j}$ is the
zero matrix of size $j\times j$ and $G_{1},G_{2},G_{3}$ are circulant
matrices of size $j\times j$. Then similar to the $3$-quasi-cyclic case  previously discussed, $F$ is a block circulant matrix and if it is of full column rank, then we can further reduce the key size by choosing a suitable $S$ such that $F$ ultimately has the column echelon form
\begin{equation}\label{4quasiF} F = \left[ \begin{array}{cc} I_j & 0_j \\ 0_j & I_j \\ F^\prime & F^{\prime\prime} \end{array} \right] \end{equation}
where the $j \times j$ matrices $F^\prime$ and $F^{\prime\prime}$ are also circulant.
\medskip

\noindent\textbf{Reduced key size}

Similarly, by considering only the first columns of each circulant matrix, we have the reduced public key size of $\frac{5}{4}nm Log (q)$ bits where $Log$ is the logarithm base 2.

\section{Security Arguments}\label{Sec:SecArg}

One advantage of using rank metric over the Hamming metric is the rapid increase in the complexity of practical attacks on finding low weight codewords as the parameters increase. This is due to the fact that finding codewords of certain weight $r$ in rank metric means finding subspaces of $\FF_{q^m}$ of dimension $r$. The number of such subspaces is given by the Gaussian binomial coefficient ${m \brack r}_q  \approx q^{(m-r)r}$. On the other hand, the number of codewords with Hamming weight $r$ is $\binom{n}{r}$ which is upper bounded by $2^n$.






\subsection{Security reduction}\label{sec:Security}
The security of the McNie cryptosystem is reduced to the RSD problem.
\begin{proof}  $ $

\begin{itemize}
\item Attacking $\mathbf{c_1} = \m G^\prime + \e$ is an instance of the RSD problem with parameters $(n,l,r)$.
\item Attacking $\mathbf{c_2} = \m F$ is an instance of the RSD with parameters $(l,l-(n-k),r)$.
\item Attacking $\mathbf{c} = (\mathbf{c_1},\mathbf{c_2}) = \m(G|F) + (\e|\mathbf{0})$ is an instance of RSD with parameters $(2n-k,l,r)$.
\end{itemize}
\end{proof}

\subsection{Practical security}
In this section we analyze the security of the proposed cryptosystem against known attacks.

\subsubsection{Direct Attacks on the message} 
An adversary can try to recover the message by directly attacking the ciphertext. The following are the best known attacks for solving the RSD problem with parameters $(n,k,r)$.




\begin{enumerate}
\item \textbf{Combinatorial attacks}
These types of attacks consider the support of a codeword  and apply the Information Set Decoding \cite{Bec} in rank metric sense. The best known strategy in \cite{GaboNew} has complexity $(n-k)^3m^3q^{r\frac{(k+1)m}{n}-m}$.

\item \textbf{Algebraic attacks}
This attack is natural for rank metric case and is most useful when $q^m$ increases. There are several types of algebraic equations settings to try to solve a multivariate system with Gr\textipa{\"o}bner basis. The best known attack in \cite{Gabo2} has complexity upper bounded by $r^3k^3q^{r\lceil \frac{(r+1)(k+1)-(n+1)}{r} \rceil}$.
\end{enumerate}

\subsubsection{Gaborit's attack \cite{PQCcomment}} 
Notice that if $\mathbf{m} = (m_1, m_2, \ldots, m_l)$ and $F$ is of full rank, then we obtain $n-k$ linear equations of the $m_i$'s from $\mathbf{c_2} = \mathbf{m}F$. Hence, all the coordinates $m_i$'s can be expressed in terms of some fixed $l-(n-k)$ coordinates.
We can then rewrite $\mathbf{c_1}$ as $\mathbf{c_1} = \mathbf{m^\prime} G^{\prime\prime} + \mathbf{e}$ where $G^{\prime\prime}$ is of dimension $l-(n-k)$. So an attacker can use the attacks above on a code of dimension $l-(n-k)$ instead of a code of dimension $l$.

This attack can be avoided by slightly modifying the encryption step. An error $\mathbf{e_2}$ is added on $\mathbf{c_2}$, i.e., $\mathbf{c_2 } = \mathbf{m}F + \mathbf{e_2}$ so that no linear relations between the coordinates $m_i$'s can be obtained. Further discussion on this modification can be found on \cite{Dual-Ouroboros}.
\subsubsection{Structural attacks}
There is a structural attack proposed by Lau, et al. in \cite{Lau} that takes advantage of the block cyclic structure of the matrices used and the assumption that $P$ is the identity matrix. In this case, the equation $FS = G^\prime H^T$ is linear and $S$ and $H$ can be solved directly. This attack can be avoided simply by using a non-identity matrix $P$.

An attacker may also attack the public matrix $F$ to obtain $H$. Since $G^\prime$ is known, an attacker can proceed with decomposing $F$ to $F=G^\prime H_0$. This would yield $q^{(n-l)(n-k)}$ possible solutions $H_0$. However, if $H$ is a low-rank parity check matrix, $H_0 = H^TS$ may not be a low rank so LRPC decoding can not be used in general.

\subsubsection{Attack on the LRPC cryptosystem \cite{Gabo3}}
This structural attack tries to attack directly the LRPC structure of the public key to recover the secret key. The fact that all the elements of the LRPC matrix $H$ belong to the same subspace $\mathcal{F}$ of rank $d$ can be used. For the dual code $D$ generated by $H$,  all the coordinates of $\mathbf{x}\in D$ belong to $\mathcal{F}$. By rewriting $\mathbf{x}=\sum_{i=1}^{n-k} a_iH_i$ for $a_i\in \FF_q$ where the $n-k$ rows of $H$ are denoted by $H_i (1\leq i\leq n-k)$ and determining $d$ $a_i$'s in $\FF_q$, it allows $\mathbf{x}$  to have $\lfloor (n-k)/d \rfloor$ coordinates of zeros since  $H$ has the weight of $d$. With a good probability, this vector $\mathbf{x}$ lies in the dual code $D$ and we may assume  the first $\lfloor (n-k)/d \rfloor$ coordinates of  $\mathbf{x}$ are all zeros without loss of generality. Now the structural attack can be done to LRPC code by choosing the subcode $D'$ of $D$ by puncturing the first  $\lfloor (n-k)/d \rfloor$ columns of $D$. Then $D'$ will be an  $[n-\lfloor (n-k)/d \rfloor, n-k-\lfloor (n-k)/d \rfloor]$ code which contains a  codeword of rank $d$.

This attack uses the structure of the LRPC matrix and the attacker only needs to find a subcode which contains at least one codeword of rank $d$. However  the computational cost of this attack is exponential. There is a result which slightly reduced its cost  using the cyclicity to decrease the number of columns of the attacked matrix. The attacker can remove columns corresponding to zeros of a small weight vector of the secret key and try to recover it. This attack is  equivalent to the attack for NTRU \cite{Hoff} and for MDPC codes \cite{Mis}.

\subsubsection{Attack on the cyclic structure \cite{Hau}}
McEliece cryptosystem based on quasi-cyclic or quasi-monoidic codes can be attacked by reducing the size of the code by adding coordinates which belong to the same orbit of the automorphism group, called the ``folding" process \cite{Fau3,Fau4}.   This process is applied to quasi-cyclic,  quasi-dyadic, alternant or Goppa codes to attack the cryptosystem for key recovery.  It was shown that the same method can be used for quasi-cyclic LRPC codes to obtain a code of much smaller size but have in its dual some low weight codewords. Then the decoding algorithm in \cite{Gabo2} can be applied to find low weight codewords in more efficient way than that of the original code.  Hauteville and Tillich show their attacks are efficient for double circulant LRPC based system, especially when the polynomial in which the folding process is generalized  can be factored. Using this method, one of the proposed parameters in \cite{Gabo1} got broken by an attack of complexity $2^{43.6}$. However this attack can be avoided by choosing the polynomial carefully. The proposed parameters in this paper are  secure against these attacks.

\subsubsection{Other attacks} 



In our proposed cryptosystem, since the public code generated by $G^\prime$ is not related to the secret code generated by $H$, attacking $G^\prime$ does not in any way expose the private code. Thus, finding a low weight codeword in the dual of $G^\prime$ is useless since $G^\prime$ is randomly chosen, and in general, not LRPC. This allows us to have freedom on choosing our low rank $d$ as it does not affect the public keys. Since LRPC decoding can be used as long as $rd \leq n-k$, we can choose $d$ small enough and $r$ high enough for increased security.


For the message attack, in general, if $G^\prime$ has a smaller error-correction capability compared to $H$, then decoding using $G$ will fail. Also, since $G^\prime$ is randomly generated, this is the rank-RSD problem. In this proposal, we can select good parameters in order to increase the decoding failure probability of using the public matrix $G^\prime$ while keeping a low decoding failure probability for the private matrix $H$. 
An attacker may also use $c_2$ to recover $\m$ by solving a system.  As mentioned in Remark 2, the proposed cryptosystem is designed to have $q^{(l-rank(F))}$ possible solutions. Although $c_2=\m F$ resembles the ciphertext from the Niederreiter cryptosystem, notice that there is no restriction on the weight of the message $\m$ so attacking the Niederreiter cryptosystem does not completely compromise our system.

\subsection{Semantic security}



In McNie, the generator matrix $G'$ and the scrambler matrix $S$ are randomly generated. Using $S$, the structure of the parity check matrix $H$ of an LRPC code is masked  to have $F$. The approach introduced in \cite{Mis} for the MDPC cryptosystem on the indistinguishability to random codes and the CCA-2 conversion in \cite{Kob} can be adapted in McNie as follows:
\begin{table}[h]
\begin{tabular}{m{1.5cm}ll}
\multicolumn{2}{l}{\bf{Notations}}&\\
 $Prep(\m)$&: &  Preprocessing to a message $\m$, such as data-compression, data-padding\\
 && and so on. Its inverse is represented as $Prep^{-1}()$.\\
$Hash(\mathbf{x})$&:&One-way hash function of an arbitrary length binary string $x$ to a \\
&& fixed length binary string. \\
$Conv(\mathbf{\bar{z}})$&:& Bijective function which converts a  vector $\mathbf{\bar{z}}$ over $\FF_{q^m}$  into the corres-\\
&& ponding  error vector $\mathbf{z}$ of length $n$ with a constant rank weight $r$. \\
&&Its inverse is  represented as $Conv^{-1}()$.\\
 $Gen(\mathbf{x})$&:&Generator of a cryptographically secure pseudo random sequences of \\
 && arbitrary length from a fixed length seed $x$.\\
$Msb_{x_1}(x_2)$&:& The left $x_1$ bits of $x_2$.\\
$Lsb_{x_1}(x_2)$&:& The right $x_1$ bits of $x_2$.\\
$Const$&:& Predetermined constant used in public.\\
$Rand$&:& Random source which generates a truly random (or computationally\\
&& indistinguishable pseudorandom) sequence.\\
$\mathcal{E}^{McNie}(x,z)$&:&Encryption of $x$ using the McNie PKC with an error vector $z$.\\
$\mathcal{D}^{McNie}(x)$&:& Decryption of $x$ using the McNie PKC.
\end{tabular}
\end{table}



\begin{table}[H]
\centering
\begin{tabular}{|rm{3.7cm}|rm{3.7cm}|}
\hline
\multicolumn{2}{|l}{\bf{~Encryption of $\bf{m}$:}} &\multicolumn{2}{|l|}{\bf{~Decryption of $\bf{c}$:}} \\
$\mathbf{r}$& $:=Rand$ & $y_5$& $:=Msb_{n-l}(\mathbf{c})$\\
 $\mathbf{\bar{m}} $& $:= Prep(\m)$ & $ z$ & $:=Lsb_{2n}(\mathbf{c})$\\
$y_1$&$ := Gen(\mathbf{r}) \oplus (\bar{\m}||Const)$ & $\mathbf{c_1}$&  $:=Msb_n(z)$\\
$y_2$& $:=\mathbf{r}\oplus Hash(y_1)$ & $\mathbf{c_2}$& $:=Lsb_n(z)$\\
 ($y_5||y_4||y_3$)& $:=(y_2||y_1)$ & $y_3$ & $:=\mathcal{D}^{McNie}(\mathbf{c_1}|| \mathbf{c_2})$\\
$\e$ & $:=Conv(y_4)$ & $\e$ & $:=y_3G'\oplus \mathbf{c_1}$\\
($\mathbf{c_1}||\mathbf{c_2}$)&$:=\mathcal{E}^{McNie}(y_3,\mathbf{e})$ & $y_4$ & :=$Conv^{-1}(\e)$\\
$ \mathbf{c}$&$:=(y_5||\mathbf{c_1}||\mathbf{c_2})$  & $(y_2||y_1)$ & $:= (y_5||y_4||y_3)$\\
return $\mathbf{c}$ & & $ \mathbf{r}$ &$:=y_2\oplus Hash(y_1)$\\
& &$\bar{\m}||Const'$ &$:=Gen(\mathbf{r})\oplus y_1$\\
&& If  & $Const'=Const$\\
& && return  $Prep^{-1}(\mathbf{\bar{m}})$\\
& & Otherwise& reject $\mathbf{c}$\\
\hline
\end{tabular}
\caption{McNie conversion }
\label{ind2}
\end{table}

The lengths of $y_3,~y_4$, and $y_5$ are as follows.
\begin{itemize}
\item $Len(y_3)=\lfloor \frac{lm}{8}\rfloor$ bytes.
\item $Len(y_4)= \lfloor \Big(\frac{r(r-1)}{2}+r(m+n-2r)\Big)/8 \rfloor$ bytes.
\item $Len(y_5)=Len(\bar{\mathbf{m}})+Len(Const)+Len(\mathbf{r})-Len(y_4)-Len(y_3)$ bytes.
\item If $Len(\bar{\mathbf{m}})+Len(Const)+Len(r)=Len(y_4)+Len(y_3)$, remove $y_5$.
\end{itemize}

Referring to \cite{Gabo1}, it is possible to use the approach in \cite{Fuj} which permits that no information is given in the case of decryption failure. This approach is used in NTRU and the MDPC code based cryptosystem as well. 
\bigskip

\section{Suggested Parameters}\label{Sec:Para}

The following table provides our suggested parameters for the given security levels.
 {\small
\begin{table}[h]
\centering 
\begin{tabular}{|c|c|c|c|c|c|c|c|c|c|}
\hline
\multirow{2}{*}{$n$ } & \multirow{2}{*}{$l$ } & \multirow{2}{*}{$k$ } & \multirow{2}{*}{$d$ } & \multirow{2}{*}{$r$ } & \multirow{2}{*}{$m$ } & \multirow{2}{*}{$q$ } & \multirow{2}{*}{failure} & Key Size & \multirow{2}{*}{security}\tabularnewline
 &  &  &  &  &   &  &  & (bytes) & \tabularnewline
\hline
\hline
120  & 80  & 80  & 3  & 8  & 53  & 2  & -23  &  795  & 128\tabularnewline
\hline
138  & 92  & 92  & 3  & 10  & 67  & 2  & -25  & 1156  & 192\tabularnewline
\hline
156 & 104  & 104  & 3  & 12  & 71  & 2  & -27  & 1385  & 256\tabularnewline
\hline 
\end{tabular}
\caption{Suggested parameters for McNie PKE using $3$-quasi-cyclic codes for 128, 192, 256-bit security levels}
\label{reduced4quasi}
\end{table} }

{\small
\begin{table}[h]
\centering
\begin{tabular}{|c|c|c|c|c|c|c|c|c|c|}
\hline
\multirow{2}{*}{$n$ } & \multirow{2}{*}{$l$ } & \multirow{2}{*}{$k$ } & \multirow{2}{*}{$d$ } & \multirow{2}{*}{$r$ } & \multirow{2}{*}{$m$ } & \multirow{2}{*}{$q$ } & \multirow{2}{*}{failure} & Key Size & \multirow{2}{*}{security}\tabularnewline
 &  &  &  &  &   &  &  & (bytes) & \tabularnewline
\hline
\hline
92  & 46  & 69  & 3  & 10  & 59  & 2  & -36 & 849  & 128\tabularnewline
\hline
112 & 56  & 84  & 3  & 13  & 67  & 2  & -38  & 1173  & 192\tabularnewline
\hline
128 & 64  & 96  & 3  & 16  & 73  & 2  & -36  & 1460  & 256\tabularnewline
\hline
\end{tabular} 
\caption{Suggested parameters for McNie PKE using $4$-quasi-cyclic codes for 128, 192, 256-bit security levels}
\label{reduced4quasi}
\end{table} }
In Table \ref{reduced4quasi}, we use an $[n,l]$ public matrix $G^\prime$ with the form given in (\ref{4quasiG}) and secret parity check matrix $H$, with the form given in (\ref{4quasiH}), for an $[n,k]$ LRPC code of weight $d$. By decryption failure we mean the probability that the LRPC decoding using $H$ will fail, expressed as a power of $2$. For example, if we use a $[72,36]$ LRPC code for the private key, i.e. a $36 \times 72$ LRPC matrix $H$, the decryption can be carried out and will fail with a very small probability of $1/2^{23}  \approx 1.19 \times 10^{-7}$.

 {\small
\begin{table}[h]
\centering
\begin{tabular}{|c|c|c|c|c|c|c|}
\hline
{Security} & \multicolumn{2}{|c|}{McNie}& {DC-LRPC }& {DC-MDPC }&{QD-Goppa}& {Goppa }     \\
\hhline{~--~~~~}
  Level & 3-quasi & 4-quasi & \cite{Gabo4} & \cite{Mis} &  \cite{Mis1} & \cite{Bern} \\
\hline
\hline
128 & \bf{6360}&\bf{6785} & 2809 & 9857 & 32768 & 1537536\\
\hline
192 & \bf{9246}&\bf{9380}& - & - &45056& 4185415 \\
\hline
256 & \bf{11076}&\bf{11680} & - & 32771 & 65536 & 7667855\\
\hline
\end{tabular}
\caption{Comparison of key sizes for McNie PKE with other code-based cryptosystems}
\label{comparison}
\end{table} }
We compare the key sizes of our proposed cryptosystem with known McEliece based cryptosystems, for different security levels. The values appearing in Table \ref{comparison} are given in bits. The values appearing in the fourth column is the McEliece cryptosystem using double-circulant LRPC codes by Gaborit, et al in \cite{Gabo1}. The fifth column uses double-circulant MDPC codes from \cite{Mis}. The last two columns use QD-Goppa codes \cite{Mis1} and Goppa codes \cite{Bern}, respectively. 

\footnotetext[2]{This is the value given in \cite{Gabo1} but was broken using the folding and projection attack in \cite{Hau}. However, there was no suggestion on new parameters.}

\section{Conclusion}
We have introduced a new public key cryptosystem that combines McEliece and Niederreiter cryptosystems. The proposed cryptosystem uses a parity check matrix of an $[n,k]$ in the private key and the generator matrix of an $[n,l]$ linear code in the public key is random in the sense that it can be freely chosen independent of the private key. Hence, the system is resistant against structural attacks on McEliece public key cryptosystems. We have also suggested several parameters for different security levels for McNie using $3$-quasi-cyclic and $4$-quasi-cyclic LRPC codes. Although we were able to obtain relatively small key sizes, this version of McNie suffers from the disadvantage of having non-zero decoding failure probability. However, McNie still seems to be a promising candidate for post-quantum cryptography. We have used LRPC code in particular in this paper but any code can be used in the McNie cryptosystem. For instance, Lau and Chan \cite{Lau} proposed a version of McNie using Gabidulin codes. They showed that Overbeck's attack \cite{Over} and other structural attacks do not work for this version of McNie. Variants of McNie using other families of codes and their cryptanalysis will be interesting research in the future.

%

\end{document}